\documentclass[10pt,preprintnumbers,aps,amssymb,nofootinbib,amsmath,superscriptaddress,notitlepage]
{revtex4-2}
\usepackage{epsfig,epsf}
\usepackage{comment}
\usepackage{bm} % puts greek and math symbols in boldface using \bm
\usepackage{color} % {\color{red} ... }
\usepackage{slashed}
\usepackage{relsize}	%rescalable math 
\usepackage{soul} %spac­ing out (let­terspac­ing), un­der­lin­ing, strik­ing out, etc.
\usepackage{hyperref}
\usepackage{tensor} %\indices{^a_b} produces the properly spaced the tensor indices 
\usepackage{yfonts} %Gotic fonts \textgoth{This: is: Gothic}
\newcommand{\beq}{\begin{equation}}
\newcommand{\beql}[1]{\begin{equation}\label{#1}}
\newcommand{\eeq}{\end{equation}}
\def\bal#1\gal{\begin{align}#1\end{align}}
\newcommand{\ball}[1]{\bal\label{#1}}
\newcommand{\eq}[1]{(\ref{#1})}
\newcommand{\fig}[1]{Fig.~\ref{#1}}
\renewcommand{\sec}[1]{Sec.~\ref{#1}}
\DeclareMathOperator{\Tr}{\mathrm{Tr}}

\renewcommand{\b}[1]{{\bm #1}} 
\newcommand{\unit}[1]{\hat {{\bm #1}}} % unit vector

\begin{document}

\title{Chiral Cherenkov radiation in the presence of a time-dependent chiral chemical potential}

\author{Kirill Tuchin}

\affiliation{
Department of Physics and Astronomy, Iowa State University, Ames, Iowa, 50011, USA}

\date{\today}

\begin{abstract}

The photon production process $f\to f+\gamma$ in the presence of a time-dependent chiral chemical potential $\mu_5$ is studied. The validity of the adiabatic approximation is demonstrated in the ultra-relativistic limit. Analytical expressions for the photon spectrum are derived for two models of the chiral magnetic conductivity $b_0\propto \mu_5$: (i) $b_0(t)= A_1+B_1\tanh(t/\tau)$ and (ii) $b_0(t)= A_2(t/\tau)/(1+t^2/\tau^2)$. It is known that for constant $b_0$, photon emission  may exhibit a resonance for one photon polarization depending on the sign of $b_0$. It is shown that in model (i), up to two resonances may occur, depending on the sign of the asymptotic values $b_0(\infty)$ and $b_0(-\infty)$. No resonances are observed in model (ii). Numerical calculations are performed using parameters relevant to quark-gluon plasma, and  the universality of the results is discussed.

\end{abstract}

\maketitle

%%%%%%%%%%%%%%%%%%%%%%%%%%%%%%%%%%%%%%%%
\section{Introduction}\label{sec:a}

One of the most intriguing features of chiral media is the chiral magnetic effect whereby electric current is induced in the direction of the external magnetic field \cite{Kharzeev:2004ey,Kharzeev:2007jp,Kharzeev:2009fn,Kharzeev:2007tn,Fukushima:2008xe}. The corresponding transport coefficient $b_0$, known as the chiral magnetic conductivity, is proportional to the chiral chemical potential $\mu_5$, which reflects the net chirality in the medium. The  chiral imbalance can have various sources. In quark-gluon plasma it may be induced by the parallel chromoelectric and chromomagnetic fields in the initial stage following a heavy-ion collision \cite{Lappi:2017skr} or by a sphaleron transition at a later stage \cite{Fukushima:2008xe}. The phenomenological manifestation of the chiral magnetic effect is electric charge separation with respect to the collision plane of the heavy ions \cite{Kharzeev:2007tn}. Recent experimental measurements have produced encouraging signs of this charge separation \cite{STAR:2025uxv}. 
Another manifestation of the chiral magnetic effect, is the chiral Cherenkov radiation, in  which a circularly polarized photon is emitted by a charges fermion \cite{Tuchin:2018sqe,Huang:2018hgk}. This process, kinematically prohibited in the empty space, becomes possible owing to the momentum supplied by the chiral magnetic current. The chiral Cherenkov radiation contributes to the parton energy loss and the jet structure, which can be directly observed in relativistic heavy-ion experiments.  

The time evolution of the chiral chemical potential is governed by three processes. The first is above-mentioned sphaleron transition, a stochastic process that generates the chiral charge.  At very high temperatures, the transition time is estimated to be significantly shorter than the lifetime of the quark-gluon plasma \cite{Arnold:1996dy,Moore:2010jd}. However, at phenomenologically relevant temperatures, these time scales are comparable. The second process is a graduate relaxation of the chiral chemical potential due to the inelastic processes breaking the chiral symmetry \cite{Hongo:2022izs}. The third process is chirality transfer to the electromagnetic field via the inverse chiral cascade \cite{Joyce:1997uy,Boyarsky:2011uy,Hirono:2015rla,Akamatsu:2013pjd,Tuchin:2014iua,Buividovich:2015jfa,Manuel:2015zpa,Kirilin:2017tdh}. These processes impart time-dependence to the chiral chemical potential, which, in turn, impacts the chiral magnetic effect, the charge separation \cite{Tuchin:2018rrw}, and the chiral Cherenkov radiation. The latter is the subject of this paper.  

The chiral Cherenkov radiation is the excitation of the electromagnetic field in a chiral medium by a charged fermion. If the time-dependence of  $b_0$ is neglected, the eigenstates of the electromagnetic field are circularly polarized monochromatic plane waves with a dispersion relation dependent on the photon polarization. The transition amplitude is then proportional to the delta function, which ensures energy conservation. However, when $b_0$ has a non-trivial time-dependence, the delta function gives way to a distribution proportional to the Fourier image of the photon amplitude. The relationship between the photon production rate and this amplitude is provided by perturbation theory. 
 
The main goal of this paper is to compute the photon production rate, given the time-dependence of $b_0$. The precise form of $b_0$ is unknown. It therefore advantageous to employ models that both provide  a qualitative description of its time-evolution, and are amenable to analytical methods. Two models are examined: (i) $b_0= A_1+B_1\tanh(t/\tau)$ and (ii) $b_0= A_2(t/\tau)/(1+t^2/\tau^2)$, where $A_1,B_1,A_2$ and $\tau>0$ are real numbers. Model (i) describes a transition between states characterized by different asymptotic values of $b_0$, whereas model (ii) describes a burst of chirality with a  finite duration. It will be shown that the models exhibit qualitatively distinct spectra, differing  from the spectrum associated with constant $b_0$.

The paper is structured as follows. In \sec{sec:b} the general formalism for computing the photon production rate at the leading order of  perturbation theory in the ultra-relativistic approximation is reviewed. The main formula for the transition probability is \eq{c9}. The photon amplitude $a(t)$ satisfies a second order differential equation \eq{b5}, which is solved for a given $b_0(t)$. Sec.~\ref{sec:c} lays out the benchmark results for the photon rate at constant $b_0$. \fig{fig:const} shows the corresponding photon spectrum. The exact solution to \eq{b5} is known only in a few relevant cases. Fortunately, all applications to relativistic heavy-ion collisions involve photons whose momenta $k$ are so large that one can use the adiabatic approximation, thereby obtaining a general expression for the field amplitude in \sec{sec:e}. The main results are derived in \sec{sec:f} and \sec{sec:g} where the photon spectra for the models (i) and (ii) are obtained and analyzed. The results are exhibited in Figs.~\ref{fig:thanh}, \ref{fig:thanh2},\ref{fig:thanh3} and \ref{fig:z}. The final section summarizes and discusses the results.

%%%%%%%%%%%%%%%%%%%%%%%%%%%%%%%%%%%%%%%%
\section{Photon radiation probability}\label{sec:b}

The electromagnetic field, in the radiation gauge $A^0=0$, $\b \nabla\cdot \b A=0$, is described by the vector potential $\b A$ obeying the equation:
\ball{b1}
\nabla^2 \b A -\partial_t^2\b A+ b_0(t)\b \nabla\times \b A=0\,.
\gal
It is assumed that the chiral magnetic conductivity $b_0$ is a given function of time unaffected by the radiation. The wave function of a photon of energy $\omega$ and momentum $\b k$ is a plane wave solution to \eq{b1}: 
\ball{b3}
\b A(t, \b r)= \frac{1}{\sqrt{V}}a(t)\b \epsilon_\lambda e^{ i\b k\cdot \b r}\,,
\gal
where the polarization vector $\b\epsilon_\lambda$,  $\lambda=\pm 1$,  satisfies the condition $i\unit k \times \b\epsilon_\lambda = \lambda \b\epsilon_\lambda$ which implies that photons must be circularly polarized.  Substituting \eq{b3} into \eq{b1} one finds that the amplitude $a(t)$ is governed by the equation
\ball{b5}
\ddot a(t)+\Omega^2(t)a(t) =0\,,
\gal
where 
\ball{b7}
\Omega^2(t)= k^2 -\lambda b_0(t)k\,.
\gal
The photon wave function will be normalized to one photon per unit volume as usual.  

Denote the energy and momentum of the incident and final fermions as $E,\b p$ and $E', \b p'$ respectively. The wave function of an incident  fermion with momentum $\b p$ and spin $s$ reads
\ball{c1}
\psi_{i} (t, \b r)= \frac{1}{\sqrt{2EV}}u_s(\b p)e^{-iEt+i \b p\cdot \b r}\,,
\gal
where $E=\sqrt{p^2+m^2}$. The wave function of the final fermion $\psi_f$ is described by a similar expression.  The photon radiation amplitude  is given by 
\ball{c3}
S_{fi}&= -ie \int \bar \psi_f\b \gamma\cdot \b A^*\psi_i d^3x dt
=-ie\frac{\bar u _{s'}(\b p')\b\gamma\cdot \b \epsilon^*_\lambda u_s(\b p)}{2V^{3/2}\sqrt{EE'}}(2\pi)^3\delta(\b p-\b p'-\b k)\tilde a^*(E-E')\,,
\gal
where I defined:
\ball{c5}
\tilde a(E-E')= \int_{-\infty}^{+\infty} e^{-i(E'-E)t}a(t)dt\,.
\gal
Given a bounded function $b_0(t)$, the convergence of the integral in \eq{c5} as assured by inserting a positive infinitesimal parameter $\epsilon$:
\bal\label{c6}
\tilde a(E-E')=\int_0^\infty e^{-i(E'-E-i\epsilon)t}a(t)dt+ \int_{-\infty}^0 e^{-i(E'-E+i\epsilon)t}a(t)dt\,.
\gal
The photon radiation probability  is then calculated as
\bal
dw &= \frac{1}{2}\sum_{\lambda,s,s'}|S_{fi}|^2\frac{V d^3p'}{(2\pi)^3}\frac{V d^3k}{(2\pi)^3}= \frac{e^2}{8(2\pi)^3 E}\sum_\lambda |\tilde a(E-E')|^2d^2k_\bot \frac{dx}{1-x} \Tr[(\slashed p+m)\slashed \epsilon^*_\lambda (\slashed p'+m)\slashed \epsilon_\lambda]\,,\label{c8}
\gal
where $x$ denotes the fraction of the incident fermion momentum carried away by the photon. Assuming that the incident fermion moves  along the $z$-axis,  $x=k_z/p_z$ and $k^2= x^2p_z^2+k_\bot^2$, where $\b k_\bot$ is the photon's transverse momentum.  The product of the photon polarization vectors can be written as:
\bal\label{ca1}
\epsilon^{\mu *}_\lambda \epsilon^\nu_\lambda = 
\left\{\begin{array}{ll}
\frac{1}{2}\left( \delta^{ij}-\frac{k^ik^j}{k^2}\right) +\frac{i\lambda}{2}\varepsilon^{ij\ell}\frac{k^\ell}{ k}\,, & \mu=i, \nu=j\,, \\
0\,, & \mu\nu=0 \,,
\end{array}\right.
\gal
where no summation over $\lambda$ is implied. 
 
In most applications the incident fermion can be regarded as ultra-relativistic meaning that $p_z\gg k_\bot,m$. In this approximation, 
\bal
E&= p_z\left( 1+ \frac{m^2}{2p_z^2}\right)\,,\label{c11}\\
E'&= (1-x)p_z\left( 1+\frac{m^2+k_\bot^2}{2(1-x)^2p_z^2}\right)\,,\label{c12}
\gal
and the differential probability \eq{c8} becomes:
\bal
\frac{dw}{d^2k_\bot dx}&= \frac{1}{8(2\pi)^3 E}\sum_\lambda |\tilde a(E-E')|^2\frac{2e^2}{x^2(1-x)^2}\left[ k_\bot^2(2-2x+x^2)+m^2 x^4\right]\,. \label{c9}
\gal
The problem of computing the photon spectrum thus reduces to determining the Fourier image \eq{c6} of the  field amplitude $a(t)$.

%%%%%%%
\section{Photon spectrum at constant $b_0$}\label{sec:c}

It is instructive to first consider the particular case of constant $b_0$  that was discussed before  \cite{Tuchin:2018sqe,Huang:2018hgk,Tuchin:2018mte,Barredo-Alamilla:2023xdt}. In this case the positive frequency solution of \eq{b5} is  
\ball{c14}
a=\frac{1}{\sqrt{2\omega}}e^{-i\omega t }\,,
\gal
where  $\omega$ is the photon energy given by 
\ball{c16}
\omega  =\sqrt{k^2+M^2}\approx xp_z\left( 1+ \frac{M^2+k_\bot^2}{2x^2p_z^2}\right)\,,
\gal
and I introduced a new parameter  $M^2= -\lambda b_0 k$. It follows then from \eq{c5} that 
\ball{c18}
\tilde a(E-E')= 2\pi \delta (\omega+E'-E)\,. 
\gal
Using \eq{c11} and \eq{c12}, the argument of the delta-function  can be written as
\ball{c20}
\omega+E'-E= \frac{k_\bot^2+M^2(1-x)+m^2x^2}{2x(1-x)p_z}\,.
\gal
The inverse of this expression is the coherence length of the photon emission: $\ell_\text{c}=1/|\omega+E'-E|$.
It signifies the distance traversed by the fermion as it radiates a photon carrying the fraction $x$ of its momentum, the transverse momentum $\b k_\bot$ and the polarization $\lambda$.
Substituting \eq{c20},\eq{c18} into \eq{c9} yields the differential photon radiation rate $d \dot w$ at constant $b_0$ \cite{Tuchin:2018sqe}:
\ball{c24}
\frac{d\dot w}{dx d^2k_\bot}= \frac{e^2}{8\pi^2 E x} \left[ \lambda E b_0 \left(1-x+\frac{x^2}{2}\right)-m^2x\right]
\delta\left( k_\bot^2-\lambda b_0 E x(1-x)+m^2x^2\right)\,.
\gal

%%%%
\begin{figure}[ht]
\begin{tabular}{cc}
      \includegraphics[width=0.45\linewidth]{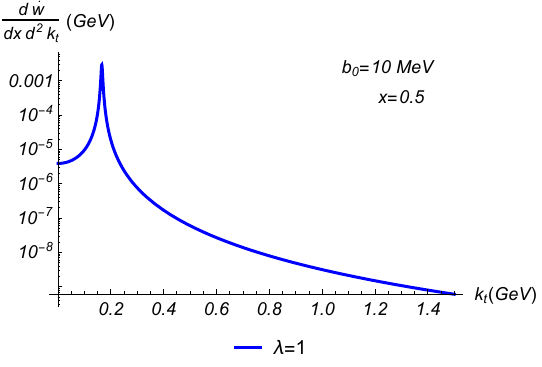} &
       \includegraphics[width=0.45\linewidth]{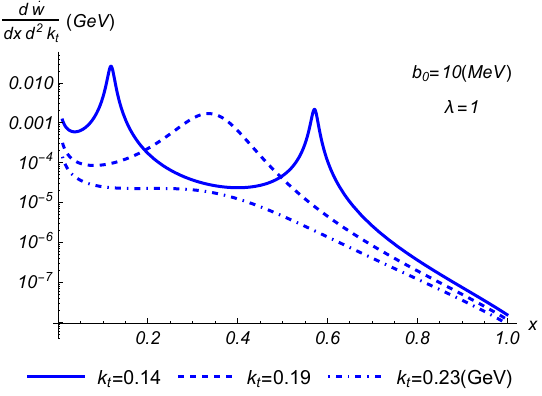} 
      \end{tabular}
  \caption{The photon spectrum  for a constant $b_0=10$~MeV. The incident fermion energy $E=20$~GeV, mass $m=0.3$~GeV and unit charge. The Lorentzian parameter $\epsilon=1$~MeV. Only the right-handed photons $\lambda=1$ are emitted.}\label{fig:const}
\end{figure}
%%%%%

The photon spectrum \eq{c24} is plotted in \fig{fig:const} for a set of parameters relevant in relativistic heavy-ion collisions. The delta-function is approximated by the Lorentzian: $\delta(x)\approx \epsilon/[\pi(\epsilon^2+x^2)]$. The peaks of the spectrum correspond to the values of $k_\bot$ and $x$ at which the argument of the delta function in \eq{c24} vanishes.  At those values the coherence length $\ell_\text{c}$ is divergent (in \fig{fig:const} and the following figures it is regulated by $\epsilon$), as it does in the classical Cherenkov radiation.  The radiated photons are right or left-handed circularly polarized if $b_0>0$ or $b_0<0$ respectively. 

Throughout the paper, the following color coding is employed: right-handed photons are plotted in blue, left-handed photons are plotted in green, and their overlap is plotted in turquoise.

%%%%%%%%%%%%%%%%%%%%%%%%%%%%%%
\section{Adiabatic approximation}\label{sec:e}

Eq.~\eq{b5} cannot be analytically solved for an arbitrary function $b_0(t)$. However, for a sufficiently slow time-variation the solution can be obtained in the adiabatic approximation.  In this section, a general formula for the amplitude $a(t)$ based on this approximation is derived. First, introduce an auxiliary function $R(t)$ defined as
\ball{e1}
a(t)=e^{iR(t)}\,.
\gal
Substituting \eq{e1} into  \eq{b5} one finds that it satisfies the equation
\ball{e3}
i\ddot R-\dot R^2+\Omega^2=0\,.
\gal
The adiabatic approximation consists in treating $R(t)$ as a slowly varying function such that 
\bal\label{e4}
|\ddot R|\ll \dot R^2\,.
\gal
To develop the adiabatic expansion, introduce a bookkeeping parameter $\alpha$ into \eq{e3}:
\bal
i\alpha\ddot R-\dot R^2+\Omega^2=0\,,\label{e5}
\gal
and substitute in it the expansion 
\bal\label{e6}
R= \sum_{j=0}^\infty \alpha^j R_j\,. 
\gal
By collecting the terms of the same order in $\alpha$  one obtains equations that govern the functions $R_0$ and $R_1$:
\bal
\dot R_0^2&= \Omega^2\,, \label{e8}\\
i\ddot R_0&= 2\dot R_0\dot R_1 \label{e9}\,.
\gal
Solving these equations and plugging the result back into \eq{e1} yields the desired result, after setting $\alpha=1$:
\bal\label{e11} 
a(t)= \frac{1}{\sqrt{2\Omega}}e^{- i\int_0^t \Omega(t')dt'}\,.
\gal
The sign in the exponent corresponds to the outgoing wave as appropriate for the emitted photon.  Expanding $\Omega$ in the ultra-relativistic approximation $k\gg b_0$:
\bal\label{e13}
\Omega(t)\approx k-\frac{\lambda b_0(t)}{2}
\gal
leads to the final result for the amplitude:
\bal\label{e15}
a(t)= \frac{1}{\sqrt{2k}}e^{- ikt +\frac{i\lambda}{2}\int_0^t b_0(t')dt'}\,.
\gal
It is valid as long as \eq{e4} is satisfied which is equivalent to the condition:
\ball{e17}
\left| \dot b_0\right|\ll k_z^2\,.
\gal

Given $b_0(t)$ one can compute the amplitude $a(t)$ using \eq{e15} and its Fourier transform \eq{c6}. The latter can be carried out analytically in a number of cases, but in general one has to rely on numerical methods. The next section deals with two examples which can be solved analytically.

%%%%%%%%%%%%%%%%%%%%%%%%%%%%%%
\section{Photon spectrum for $b_0(t)=A_1+ B_1\tanh\frac{t}{\tau}$}\label{sec:f}

In this section I consider the following model for the time evolution of the chiral magnetic conductivity:
\bal\label{d1}
b_0(t)=A_1+ B_1\tanh\frac{t}{\tau}\,.
\gal
The initial and final values $b_0(t)$ are  $b_0(-\infty)=A_1-B_1$ and $b_0(\infty)= A_1+B_1$ respectively. The transition time between them is controlled by the parameter $\tau$.

%%%%%%%%%%%%
\subsection{Exact solution for $a(t)$}\label{sec:f1}
 
Substituting \eq{d1} into \eq{b7} one obtains the function $\Omega(t)$:
\ball{d3}
\Omega^2(t)= k^2-\lambda k A-\lambda k B \tanh \frac{t}{\tau}\,.
\gal
The exact solution of \eq{b5} with $\Omega$ given by \eq{d3} was obtained in \cite{Bernard:1977reg}. It is most readily expressed in terms of 
the asymptotic values of $\Omega$ as follows:
\bal
\Omega_\text{in}&=\sqrt{k^2-\lambda k (A_1-B_1)}\,,\label{d5}\\
\Omega_\text{out}&=\sqrt{k^2-\lambda k (A_1+B_1)}\,,\label{d6}\\
\Omega_\pm & = \frac{1}{2}(\Omega_\text{out}\pm \Omega_\text{in})\,. \label{d7}
\gal
In the ultra-relativistic approximation, expand \eq{d5},\eq{d6} and \eq{d7} at large $k_z$:
\bal
\Omega_\text{in}&= k_z-\frac{\lambda}{2}(A_1-B_1)+\mathcal{O}(k_z^{-1}) \,,\label{da1}\\
\Omega_\text{out}&=k_z-\frac{\lambda}{2}(A_1+B_1)+\mathcal{O}(k_z^{-1})\,,\label{da2}\\
\Omega_+&= k_z -\frac{\lambda A_1}{2}\,,\label{da4}\\
\Omega_-&= -\frac{\lambda B_1}{2}\,.\label{da5}
\gal
The solution to \eq{b5} describing the emitted photon, namely the one that has positive frequency at the future infinity reads:
\bal
a(t)&= \frac{1}{\sqrt{2\Omega_\text{out}}}e^{-i\Omega_+ t-i\Omega_-\tau \ln\left[2\cosh\frac{t}{\tau}\right]}
{_2}F_1\left(1+i\Omega_-\tau, i\Omega_-\tau;1+i\Omega_\text{out}\tau;\frac{1}{2}\left(1-\tanh \frac{t}{\tau}\right)\right)\,,\label{d9}\\
&\approx\frac{1}{\sqrt{2 k_z}}e^{-i\left( k_z -\frac{\lambda A_1}{2}\right) t+i\frac{\lambda B_1}{2}\tau \ln\left[2\cosh\frac{t}{\tau}\right]}
{_2}F_1\left(1-\frac{i\lambda B_1\tau}{2}, -\frac{i\lambda B_1\tau}{2};1+ik_z\tau;\frac{1}{2}\left(1-\tanh \frac{t}{\tau}\right)\right)\label{d10}\,.
\gal
Using Euler's hypergeometric transformation: ${_2}F_1(\alpha,\beta;\gamma;z)=(1-z)^{\gamma-\alpha-\beta}{_2}F_1(\gamma-\alpha,\gamma-\beta;\gamma;z)$ and assuming, consistently with the ultra-relativistic approximation, that $k_z\tau\gg 1$ and $k_z\gg |B_1|$, so that $\gamma\gg \alpha,\beta$, it emerges that the hypergeometric function in \eq{d10} reduces to unity.

%%%%%%%%
\subsection{Adiabatic approximation}\label{sec:f2}

The adiabatic approximation \eq{e15} precisely reproduces the exact solution \eq{d10} which can be seen by  substituting \eq{d1} into \eq{e15} and replacing $k\approx k_z$:
\ball{fda7}
a(t)= \frac{1}{\sqrt{2 k_z}}e^{-i\left( k_z -\frac{\lambda A_1}{2}\right) t+i\frac{\lambda B_1}{2}\tau \ln\left[2\cosh\frac{t}{\tau}\right]}\,.
\gal
According to \eq{e17} this formula is valid when the parameters satisfy $k_z^2\tau\gg |B_1|$. Thus the adiabatic approximation accurately describes the time-evolution of the electromagnetic field in the ultra-relativistic limit which spares one seeking exact solutions to \eq{b5}.

To calculate the Fourier image of the amplitude using \eq{c6},  cast it in the form:
\bal\label{fd16}
\tilde a(E-E')&=\frac{2}{\sqrt{2k_z}}\int_0^\infty  \cos\left[\left(E-E'-k_z+\frac{\lambda A_1}{2}\right)t\right] e^{ -\left(i\frac{\lambda B_1\tau}{2}+\epsilon\right) \ln\left[2\cosh\frac{t}{\tau}\right]} dt\,.
\gal
I used the fact that at $t\to \infty$, the infinitesimal term in the exponent becomes $-\epsilon t/\tau$ as prescribed by \eq{c6} (up to the non-essential factor $1/\tau$ that simplify redefines $\epsilon$).  Employing \eq{da1}--\eq{da5},  \eq{fd16} may be written as:
\bal\label{fd16-A}
\tilde a(E-E')&=\frac{2}{\sqrt{2\Omega_\text{out}}}
\int_0^\infty  \cos\left[(E-E'-\Omega_+)t\right] e^{ -(i\Omega_-\tau+\epsilon) \ln\left[2\cosh\frac{t}{\tau}\right]} dt\,.
\gal
Performing the Fourier transform using Eq.~1.9(5) of \cite{bateman1954tables} yields: 
\bal
\tilde a(E-E')=&\frac{1}{\sqrt{2\Omega_\text{out}}}
\frac{2^{i\Omega_-\tau-1}\tau}{\Gamma(i\Omega_-\tau)} \Gamma\left(\frac{i\tau}{2}(E-E'-\Omega_\text{in}-i\epsilon)\right)
\Gamma\left(\frac{i\tau}{2}(-E+E'+\Omega_\text{out}-i\epsilon)\right)\,.\label{fd19}
\gal
Substituting \eq{fd19} into  \eq{c9} furnishes the photon spectrum:
\bal
\frac{dw}{dx d^2k_\bot}= &\frac{1}{16(2\pi)^3 E^2}\sum_\lambda \frac{2e^2}{x^3(1-x)^2}\left[ k_\bot^2(2-2x+x^2)+m^2 x^4\right]  \frac{\tau^2}{4|\Gamma(i\Omega_-\tau)|^2}\nonumber\\
&\times 
\left|\Gamma\left(\frac{i\tau}{2}(E-E'-\Omega_\text{in}-i\epsilon)\right)
\Gamma\left(\frac{i\tau}{2}(-E+E'+\Omega_\text{out}-i\epsilon)\right)\right|^2 \,.\label{fd22}
\gal
The explicit dependence of the $\Gamma$-function arguments on photon momentum is:
\bal
E'-E+\Omega_\text{in}&= \frac{k_\bot^2 -\lambda x (1-x) p_z (A_1-B_1)+m^2x^2}{2x(1-x)p_z}\,,  \label{fd25}\\
E'-E+\Omega_\text{out}&= \frac{k_\bot^2 -\lambda x(1-x)p_z(A_1+B_1)+m^2x^2}{2x(1-x)p_z}\,.  \label{fd26}
\gal
These expressions vanish at certain values of $k_\bot$ and $x$, which correspond to the resonances in the spectrum. The height and width of these resonances are regulated by the parameter $\epsilon$.

%%%%
\begin{figure}[ht]
\begin{tabular}{cc}
      \includegraphics[width=0.45\linewidth]{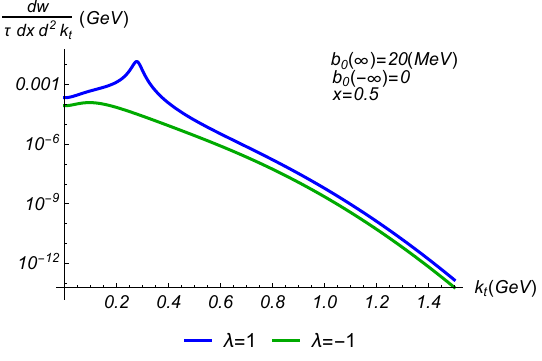} &
       \includegraphics[width=0.45\linewidth]{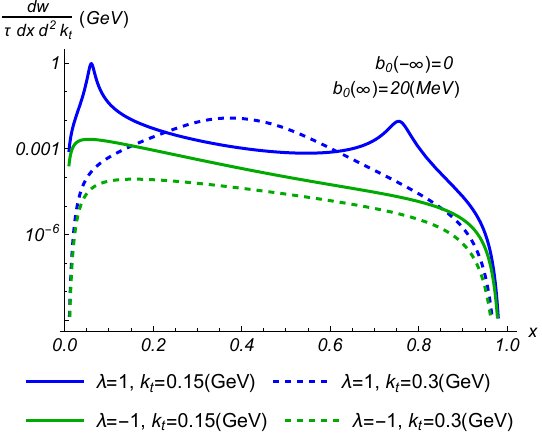} \
      \end{tabular}
  \caption{The photon spectrum for $b_0(t)=A_1+ B_1\tanh\frac{t}{\tau}$ with $A_1=B_1=10$~MeV and $\tau=25$~GeV$^{-1}\approx$ 5~fm/c.  The incident fermion energy $E=20$~GeV, mass $m=0.3$~GeV, and unit charge.  }\label{fig:thanh}
\end{figure}
%%%%%

%%%%
\begin{figure}[ht]
\begin{tabular}{cc}
      \includegraphics[width=0.45\linewidth]{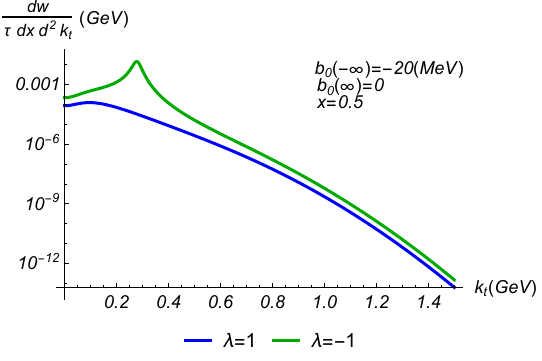} &
       \includegraphics[width=0.45\linewidth]{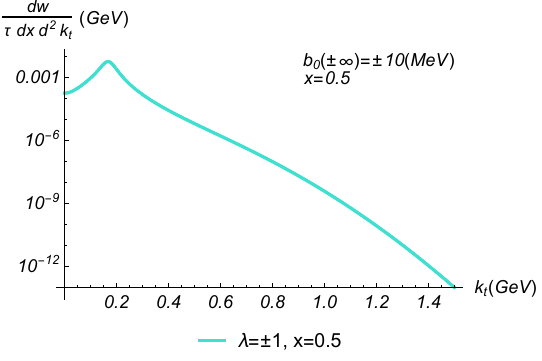} \
      \end{tabular}
  \caption{The photon spectrum for $b_0(t)=A_1+ B_1\tanh\frac{t}{\tau}$ with $A_1=-B_1=-10$~MeV (left),  $A_1=0$, $B_1=10$~MeV (right) and $\tau=25$~GeV$^{-1}\approx$ 5~fm/c. The incident fermion energy $E=20$~GeV, mass $m=0.3$~GeV, and unit  charge. The turquoise color indicates that the spectrum is independent of the photon's polarization. }\label{fig:thanh2}
\end{figure}
%%%%%

%%%%%%
\begin{figure}[ht]
       \includegraphics[width=0.45\linewidth]{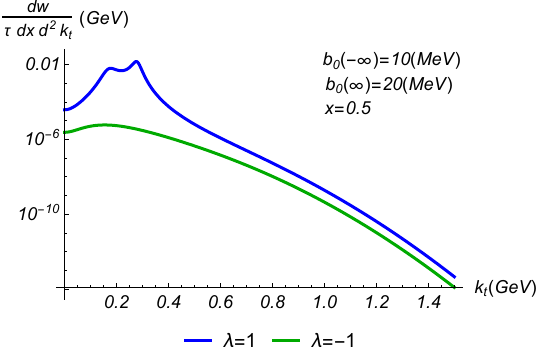} 
  \caption{The photon spectrum for $b_0(t)=A_1+ B_1\tanh\frac{t}{\tau}$ with $A_1=15$~MeV, $B_1=5$~MeV  $\tau=25$~GeV$^{-1}\approx$ 5~fm/c. The incident fermion energy $E=20$~GeV, mass $m=0.3$~GeV, and unit  charge. }
\label{fig:thanh3}
\end{figure}
%%%%%

 To plot the photon spectrum, select the values of the parameters relevant to the phenomenology of quark-gluon plasma produced in relativistic heavy-ion collisions. The chiral magnetic conductivity is estimated to be within the range of  $10-100$ MeV \cite{Kharzeev:2009pj,Gynther:2010ed,Yee:2009vw,Li:2016tel}. Assuming that the time-dependence of $b_0$ is predominantly  due to sphaleron transitions, the typical value of $\tau$ is $1/\alpha_s^2T$ \cite{Arnold:1996dy}, in the range of $1-10$~fm, shorter that the plasma's lifetime.

Figs.~\ref{fig:thanh},\ref{fig:thanh2} and \ref{fig:thanh3} present the photon spectrum for typical parameters. They describe the photon spectrum for various values of  $b_0(\infty)$ and $b_0(-\infty)$.  The resonances correspond to vanishing of expressions \eq{fd25} and \eq{fd26} which occurs only if  $\lambda(A_1-B_1)= \lambda b_0(-\infty)>0$ and $\lambda(A_1+B_1)=\lambda b_0(\infty)>0$ respectively. The positions of the resonances satisfy the equations
\bal
&k_\bot^2 -\lambda x (1-x) E (A_1\mp B_1)+m^2x^2=0\,.\label{fd30}
\gal
Clearly, a particular polarization is enhanced depending on the signs of asymptotic values $b_0(\pm \infty)$.  When one of those values vanishes, only single resonance appears in the $k_\bot$ spectrum, see \fig{fig:thanh} and \fig{fig:thanh2} (left). The two polarizations overlap in  \fig{fig:thanh2}(right) because $\lambda b_0(-\infty)= -\lambda b_0(\infty)$. In \fig{fig:thanh3} one can see two distinct resonances in the right-handed photon spectrum because both asymptotic values are positive. There are twice as many resonances   in the $x$-spectrum as compared to the $k_\bot$-spectrum  as is evident from \eq{fd25} and \eq{fd26}. This is seen in \fig{fig:thanh}(right). 

 The resonant value of $k_\bot$ at a given $x$ is
 \bal\label{fd33}
 K_\bot =\sqrt{\lambda x (1-x) E (A_1\mp B_1)-m^2x^2}\,,
 \gal
provided that 
\bal\label{fd34}
0<x\le \frac{\lambda E (A_1\mp B_1)}{\lambda E (A_1\mp B_1)+m^2}\,,
\gal
while the resonant values of $x$ at a given $k_\bot$ are:
 \bal\label{fd37}
 X_\alpha= \frac{\lambda E(A_1\mp B_1)+(-1)^\alpha \sqrt{E^2(A_1\mp B_1)^2-4\left[m^2+\lambda E (A_1\mp B_1)\right]k_\bot^2}}{2\left[ m^2+\lambda E (A_1\mp B_1)\right]}\,,\quad \alpha=1,2\,,
 \gal 
provided that 
\bal\label{fd38}
0<k_\bot^2\le \frac{E^2(A_1\mp B_1)^2}{4\left[ m^2+\lambda E (A_1\mp B_1)\right]}\,.
\gal

%%%%
\subsection{Sudden transition}

The adiabatic approximation does not require $\tau$ to be very large, as was shown in the previous subsection. In fact, it can be rather small as long as the condition $k_z^2\tau\gg |B_1|$ is fulfilled. In particular, $\tau$ can be much smaller than the coherence length $\ell_c$ defined in \sec{sec:c} for constant $b_0$. In our case, $1/\tau$ must be much larger than either of two expressions \eq{fd25} and \eq{fd26}. This condition is certainly satisfied near the resonances. It also holds for not too large values of $k_\bot$. One can refer to such a scenario as a sudden transition.  Taking the limit $\tau\to 0$ in the exponent of \eq{fda7}  yields
\ball{dc10}
a(t) =\frac{1}{\sqrt{2\Omega_\text{out}}}\left(  e^{-i\Omega_\text{out}t}\theta(t) +  e^{-i\Omega_\text{in}t}\theta (-t)\right)\,.
\gal
Its Fourier-transform \eq{c5} is
\bal
\tilde a(E-E')&= \frac{1}{\sqrt{2\Omega_\text{out}}}\left\{ \int_0^\infty e^{i(E-E'-\Omega_\text{out}+i\epsilon)t}dt
+ \int^0_{-\infty} e^{i(E-E'-\Omega_\text{in}-i\epsilon)t}dt \right\}\label{dc11}
\\
&= \frac{i}{\sqrt{2\Omega_\text{out}}}\left\{ \frac{1}{E-E'-\Omega_\text{out}+i\epsilon}-\frac{1}{E-E'-\Omega_\text{in}-i\epsilon}\right\}\,.\label{dc12}
\gal
In particular, if $b_0$ is constant, then $B_1=0$,  $\Omega_\text{out}= \Omega_\text{in}$ and \eq{dc12} reduces to \eq{c18} as can be seen by using the formula 
\ball{dc14}
\frac{1}{x\pm i\epsilon}= \mp i\pi \delta(x)+ P\frac{1}{x}\,.
\gal

%%%%%%%%%%%%%%%%%%%%%%%%%%%%%%%%%%
\section{Photon spectrum for $b(t)= A_2(t/\tau)/(1+t^2/\tau^2)$} \label{sec:g}

In this section, consider a scenario in which $b_0$ evolves in such a way that its value at distant past and future vanish. It can be described by the following model:
\bal\label{e25}
b(t)= \frac{A_2t}{\tau\left(1+\frac{t^2}{\tau^2}\right)}\,.
\gal
Unlike the previous model, the exact solution to equation \eq{b5} with \eq{e25} is not known. I will therefore employ the adiabatic approximation that was developed in \sec{sec:e} and successfully used in the previous section. Accordingly, plugging \eq{e25} into \eq{e15} yields
\ball{e27}
a(t)= \frac{1}{\sqrt{2k}}e^{-ikt}\left( 1+\frac{t^2}{\tau^2}\right)^\frac{i\lambda \tau A_2}{4}\,.
\gal
Its Fourier transform is given by 
\bal
\tilde a(E-E')=& \frac{1}{\sqrt{2k}}\int_0^\infty e^{-i(E'-E+k-i\epsilon)t}\left( 1+\frac{t^2}{\tau^2}\right)^\frac{i\lambda \tau A_2}{4}dt +\frac{1}{\sqrt{2k}}\int_{-\infty}^0 e^{-i(E'-E+k+i\epsilon)t}\left( 1+\frac{t^2}{\tau^2}\right)^\frac{i\lambda \tau A_2}{4}dt
\label{e29}\\
=& 
\frac{2}{\sqrt{2k}}\int_0^\infty \left( 1+\frac{t^2}{\tau^2}\right)^\frac{i\lambda \tau A_2}{4}\cos\left[ (E'-E+k)t\right]e^{-\epsilon t}dt\,.
\gal
This integral can be taken using Mathematica 14.2 package and the result expressed in terms of the function:
\ball{e31}
F_\nu(x)= x^{\frac{1}{2}+i\nu}\left[\nu J_{-\frac{1}{2}-i\nu}( x^*)\Gamma(-i\nu)- J_{\frac{1}{2}+i\nu}( x^*)\Gamma(1-i\nu)\sinh(\pi \nu)+\nu \cosh(\pi \nu)\Gamma(-i\nu)\b H_{\frac{1}{2}+i\nu}(x^*)\right]\,,
\gal
where $\nu = \frac{\lambda \tau A_2}{4}$ and $\b H$ is Struve's function. Thus,
\ball{e33}
\tilde a(E-E')= \frac{\pi^{\frac{3}{2}}2^{-\frac{1}{2}+i\nu}}{\sqrt{2k}\,\tau^{2i\nu}}\frac{(y^2+\epsilon^2)^{-\frac{1}{2}-i\nu}}{\cosh(\pi\nu)\sinh(\pi\nu)\Gamma(1-i\nu)\Gamma(-i\nu)}\left\{ F_\nu\left[(iy+\epsilon)\tau\right]+F_\nu\left[(-iy+\epsilon)\tau\right]\right\}\,,
\gal
where  
\bal\label{e39}
y&= E'-E+k=\frac{k_\bot^2+m^2x^2}{2x(1-x)p_z}\,.
\gal
In the sudden transition limit $\tau\to 0$ \eq{e33} vanishes. Indeed,
\ball{e35}
\lim_{\tau\to 0}\tilde a(E-E')=\frac{i}{\sqrt{2k}}\left( \frac{1}{E'-E+k+i\epsilon}-\frac{1}{E'-E+k-i\epsilon}\right) =\frac{1}{\sqrt{2k}}2\pi \delta(y) =0\,,
\gal
because $y>0$. Substituting \eq{e33}  into \eq{c9} furnishes the photon spectrum: 
 \bal
\frac{dw}{dx d^2k_\bot}= &\frac{1}{16(2\pi)^3 E^2}\sum_\lambda \frac{2e^2}{x^3(1-x)^2}\left[ k_\bot^2(2-2x+x^2)+m^2 x^4\right]  \nonumber\\
&\times 
\frac{\pi\tau}{2\cosh^2(\pi\nu)}\frac{\left| F_\nu\left[(iy+\epsilon)\tau\right]+F_\nu\left[(-iy+\epsilon)\tau\right]\right|^2}{y^2+\epsilon^2} \,.\label{e37}
\gal

\begin{figure}[ht]
\begin{tabular}{cc}
      \includegraphics[width=0.45\linewidth]{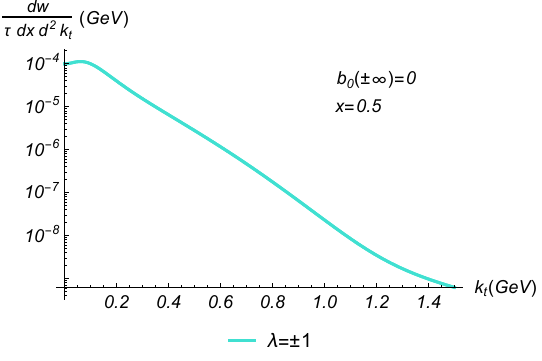} &
       \includegraphics[width=0.45\linewidth]{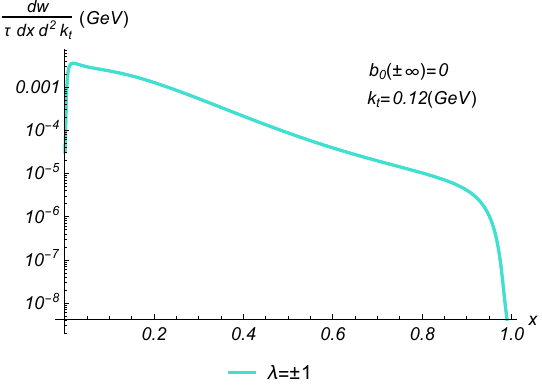} \
      \end{tabular}
  \caption{The photon spectrum for $b_0(t)= A(t/\tau)/(1+t^2/\tau^2)$  with $A_2=10$~MeV and $\tau=25$~GeV$^{-1}\approx$ 5~fm/c. The incident fermion energy $E=20$~GeV, mass $m=0.3$~GeV, and unit  charge. The spectrum does not depend on the photon's polarization, which is reflected in turquoise color, and does not exhibit any resonances.}\label{fig:z}
\end{figure}

The numerical result representing the parameters of the quark-gluon plasma is shown in \fig{fig:z}. The most conspicuous feature of the spectrum is the equality of the photon polarizations indicating that  the $\lambda$-dependence cancels out. This happens because $|F_\nu + F_{-\nu}|^2$ is an even function of $\nu$ and, consequently, of $\lambda$. Also, it is noteworthy that the resonances are absent, which aligns with the observations in the preceding section, where it was observed that the emergence of the resonances is contingent upon finite values of at least one of the asymptotics of $b_0$.
%%%%
%%%%%

%%%%%%%%%%%%%%%%%%%
\section{Summary}\label{sec:s}

The chiral Cherenkov mechanism of photon radiation was investigated using two models of the time-evolution of the chiral magnetic conductivity $b_0(t)$. The first model $b_0(t)= A_1+B_1\tanh(t/\tau)$ describes a transition of $b_0$ between two asymptotically constant values $b_0(\pm \infty)$. This mechanism allows the production of only circularly polarized photons. In general, the cross sections for the two polarizations are significantly different. The main feature of the $k_\bot$-spectrum is the potential emergence of one or two resonances. The necessary conditions of the resonance emergence are the fulfillment of either or both of the following conditions: $\lambda b_0(\infty)>0$ and $\lambda b_0(-\infty)>0$.  The positions of the resonances are given by \eq{fd33}--\eq{fd38}. The typical spectra are presented in Figs.~\ref{fig:thanh},\ref{fig:thanh2} and \ref{fig:thanh3}. In contrast, only a single resonance may appear when $b_0$ is constant as illustrated  in \fig{fig:const}.

The second model represents a pulse of chirality with a finite duration: $b_0(t)= A_2(t/\tau)/(1+t^2/\tau^2)$. In this case, there are no resonances, and the two polarizations have identical spectra. 

The two models investigated in this paper suggest that the time-dependence of the chiral chemical potential diminishes \emph{the average} polarization of the chiral Cherenkov radiation as compared to the constant $b_0$ case. The chirial Cherenkov resonance, appearing in the figures as a peak at a certain $k_\bot$ (or two peaks at the corresponding $x$'s), gives the largest contribution to the rate and  emerges if $b_0(\pm \infty)\neq 0$. It also determines the average polarization. The mathematical structure of the Fourier image of the amplitude $a(t)$ suggests that these conclusions may hold universally. Indeed, the parameter $\epsilon$ that controls the convergence of the time-integral  at $t\to \pm \infty$ in \eq{c6} also sets the resonance height, which is proportional to $1/\epsilon$. The resonance appears in any model in which $b_0$ is finite at either remote past or future. 

Our calculations significantly relied on the adiabatic approximation developed in \sec{sec:e}. This approximation enabled us to analytically compute the photon amplitude and its Fourier transform. It holds when $ k_z^2\gg |b_0|/\tau$. For $b_0$ in the range of 10-100~MeV \cite{Kharzeev:2009pj,Gynther:2010ed,Yee:2009vw,Li:2016tel} and $\tau$ in the range of $1-10$~fm, \cite{Arnold:1996dy,Moore:2010jd} it implies conservatively that  $k\gg 100$~MeV. The validity of the adiabatic approximation in a wide range of photon energies further indicates that our main results---the qualitative structure of the photon spectrum---are not model-dependent. Nevertheless, further investigation of the model space, potentially employing numerical methods, would be of great interest.

I implicitly assumed that the medium lifetime and spatial dimensions are much larger than any time or distance scale associated  with sphaleron transitions and the process $f\to f+\gamma$ . This assumption might be in tension with the relativistic heavy-ion phenomenology.  It would be interesting to investigate how the finite dimensions of the realistic quark-gluon plasma impact the results presented in this paper. 
 
Our analysis focused exclusively on applications to the quark-gluon plasma. Nevertheless, the developed formalism is equally applicable to describe photon radiation emanating from cosmic rays in the background of the axion field\cite{Saveliev:2024whq} or by electrons in Weyl semimetals \cite{Hansen:2024kvc}.

%%%%%%%%%%%%%%%%%%%%%%%%%%%%%%%%
\acknowledgments
%I  am grateful to ... for many fruitful discussions of related problems. 
%We thank ... for helpful communications/correspondance.
This work  was supported in part by the U.S. Department of Energy under Grant No.\ DE-SC0023692.

%%%%%%%%%%%%%%%%%%%%%%%%%%%%%%
\bibliography{anom-biblio}

%% appendix 
%\appendix
%\section{}\label{appA}

\end{document}